\documentstyle[12pt]{article}
\newcommand{\be}{\begin{equation}}
\newcommand{\ee}{\end{equation}}
\begin{document}

\vspace*{4cm}

\begin{center}
\large{\bf HYBRID EXCITATIONS OF THE QCD  STRING WITH QUARKS}\\

\vspace{1cm}

{\bf Yu.S.Kalashnikova\footnote{E-mail: yulia@vxitep.itep.ru},
Yu.B.Yufryakov\footnote{E-mail: yufryakov@vxitep.itep.ru}} \\

\vspace{0.5cm}

{\small Institute of Theoretical and Experimental Physics\\
117259, Russia, Moscow, B.Cheremushkinskaya 25.}

\end{center}
\vspace{1cm}
\begin{abstract}

The model for constituent glue is presented starting from the
perturbation theory in nonperturbative QCD vacuum. Green function is
constructed for a system containing $q\bar{q}$ pair and gluon
propagating in the vacuum background field, and Hamiltonian approach
to the problem is formulated. The masses of lowest $q\bar{q}g$
hybrids with light quarks are evaluated, and implication of results
for the spectroscopy of exotic states is discussed.

 \end{abstract}

 There is no doubts
now that the gluonic degrees of freedom are not only present in the
QCD Lagrangian, but should also exhibit themselves at the constituent
level in such a way that the pure gluonium and hybrid states should
exist. Experimental candidates for such exotic states appear from
time to time, causing quite natural excitement in the situation when
the  standard $q\bar{q}$ nonets are already overpopulated. On the
other hand, the state-of-art of the QCD in the strong coupling
nonperturbative regime still does not allow to predict unambigiously
the mass scale of the non-$q\bar{q}$ states.

Estimates for the masses of the hybrid $q\bar{q}g$ mesons were
carried out within the various QCD--inspired approaches, such as bag
models [1], constituent gluon models [2], flux-tube model [3,4], and
QCD sum rules [5], with rather controversial results. For example,
the predictions for the masses of lightest hybrids with $u$ and $d$
quarks vary from 1.5 GeV to 2.5 GeV. Having in mind that model
approaches usually contain both theoretical and parameter
uncertainties, we claim that more direct QCD--motivated treatment is
desirable to handle the problem.

In the present letter we study the $q\bar{q} g$ system in the
framework of Vacuum Background Correlator method [6], constructing the
hadronic Green functions starting from the QCD Lagrangian, with the
dynamics of interaction defined by the averages of corresponding
Wilson loop operators. It may be shown [6,7] that under reasonable
assumptions on the behaviour of background vacuum correlators the
area law asymptotics for Wilson loop averages  at large distances
appears, giving  rise to the string--type interaction of
constituents. The $q\bar{q}$ system with this interaction was
described in [8], and the picture
of hybrids in this approach was first considered in [9].

We introduce a model for constituent glue starting from the
perturbation theory expansion in nonperturbative QCD vacuum [10]. As
it is usually done in the background field formalism, we split the
gluonic field $A_{\mu}$ into the background part $B_{\mu}$ and the
perturbation $a_{\mu}$ over background:
\be
A_{\mu}=B_{\mu}+a_{\mu}.
\ee
Ascribing the inhomogeneous part of gauge transformation to the field
$B_{\mu}$ we can form gauge invariant states with the field
$a_{\mu}$ involved. One--gluon hybrid is represented as
\be
\Psi(x_q,x_{\bar{q}},x_g)= \bar{\psi}_{\alpha}(x_{\bar{q}})
\Phi^{\alpha}_{\beta}(x_{\bar{q}},x_g)a^{\beta}_{\gamma}(x_g)
\Phi^{\gamma}_{\delta}
(x_g,x_q)\psi^{\delta}(x_q),
\ee
where $\alpha...\delta$ are the colour indices in the fundamental
representation, spin indices  are omitted, $\psi (\bar{\psi})$ stands
for the quark (antiquark) field,
$a_{\gamma}^{\beta}=(\lambda_a)^{\beta}_{\gamma} a_a$, and parallel
transporter $\Phi$ contains only background field:
\be
\Phi^{\alpha}_{\beta}(x,y)=P exp\int^x_yB_{\mu\beta}^{\alpha}
dz_{\mu}
\ee
The evolution of the state (2) is described by the Green function
\be
G(x_qx_{\bar{x}}x_g,y_qy_{\bar{q}}y_g)=
<\Psi(y_q,y_{\bar{q}},y_g)\Psi^+(x_q,x_{\bar{q}},x_g)>_B,
\ee
where the averaging over background field configurations is
implied.

The Green function for the field $a_{\mu}$ propagating in the
given background field $B_{\mu}$ may be written in the Eucledian
space--time as [10]:
\be
G^{-1}_{\mu\nu}=D^2\delta_{\mu\nu}-g
F_{\nu\mu}-D_{\nu}D_{\mu}+\frac{1}{\xi} D_{\mu}D_{\nu}= \ee $$
=M_{\mu\nu}+\frac{1}{\xi}D_{\mu}D_{\nu},
$$
with the background gauge fixing term $G^a=(D_{\mu}a_{\mu})^a$, where
$ D_{\lambda}^{ca}=\partial_{\lambda}\delta^{ca}+gf^{cba}B^b_{\lambda}$. If
the background field satisfies the equation of motion $D_{\mu}F_{\mu\nu}=0$,
then one  has $M_{\mu\nu}D_{\nu}=0$, and the Green function (5) may be
rewritten as
\be
G_{\mu\nu}=(\delta_{\mu\nu}+D_{\mu}(\frac{1}{D^2})D_{\lambda}\cdot
(\xi-1))(D^2-2gF)^{-1}_{\lambda\nu}.
\ee
The choice $\xi=0$ corresponds to the Landau gauge, in which the Green
function (6) contains explicitly the projector $P_{\mu\lambda} $ onto
transverse states:\footnote{The condition $D_{\mu}F_{\mu\nu}=0$ is claimed
to be not necessary [11] and may be removed using the Coulomb background
gauge, which has an additional advantage: in this gauge the ghosts
can be decoupled explicitly.} \be
P_{\mu\lambda}=\delta_{\mu\lambda}-D_{\mu}(\frac{1}{D^2})D_{\lambda}
\ee

The next step is to use the Feynman--Schwinger representation [12] of
Green function (4) to define the effective action for the $q\bar{q}g$
system. Here we present the simplified version of the model,
 neglecting
spin degrees of freedom and reducing the problem to the  effective scalar
one; this approximation corresponds to the omitting the projector in the
gluon Green function (6) and colour magnetic interaction (term
proportional to $gF$). In a similar way we omit the spin dependence
in the quark Green function, assuming
\be
G_q=(D^2-m^2_q)^{-1}
\ee

 Within this
approximation the Feynman--Schwinger representation for the Green
function (4) takes the form
\be
G(x_qx_{\bar{q}}x_g; y_qy_{\bar{q}}y_g)=
\ee
$$
\int^{\infty}_0ds\int^{\infty}_0d\bar{s}\int^{\infty}_0dS\int
DzD\bar{z}DZ exp(-
 {\cal{K}}_q-{\cal{K}}_{\bar{q}}-{\cal{K}}_g)<{\cal{W}}>_B,
 $$
  where
 $${\cal{K}}_q=m^2_qs+\frac{1}{4}\int^s_0\dot{z}^2(\tau)d\tau,~~
 {\cal{K}}_{\bar{q}}=m^2_{\bar{q}}\bar{s}+\frac{1}{4}\int^{\bar{s}}_0
\dot{\bar{z}}^2(\tau)d\tau,
 ~~{\cal{K}}_g=\frac{1}{4}\int^S_0\dot{Z}^2(\tau)d\tau,$$
 boundary conditions read
 $$z(0)=x_q,~~\bar{z}(0)=x_{\bar{q}},~~Z(0)=x_g,$$
 $$z(s)=y_q,~~\bar{z}(\bar{s})=y_{\bar{q}},~~Z(S)=y_g,$$
 and all the dependence on the gluonic fields $B$ is contained in the
 Wilson loop operator
 \be
 {\cal{W}}=\lambda_a\Phi_{\Gamma_q}(y_q,x_q)\lambda_b
 \Phi_{\Gamma_{\bar{q}}}(x_{\bar{q}},
 y_{\bar{q}})\Phi_{ab,\Gamma_g}(y_g,x_g),
 \ee
 the contours $\Gamma_q,\Gamma_{\bar{q}}$ and $\Gamma_g$ run over the
 trajectories of the quarks and the gluon correspondingly (see Fig.),
 $\Phi_{ab}$ is the transporter in the adjoint representation.

 The Wilson loop configuration (10) can be disentangled using the
 relation between ordered exponents along the gluon path $\Gamma_g$
 in the adjoint and fundamental representation:
 \be
 \frac{1}{2}
 \Phi_{ab,\Gamma_g}(x,y)=tr(\lambda_a\Phi_{\Gamma_g}(x,y)
 \lambda_b\Phi_{\bar{\Gamma}_g} (y,x)),
 \ee
 (path $\bar{\Gamma}_g$ is directed oppositely to $\Gamma_g$), with
 the result
 \be
 <{\cal{W}}>_B=\frac{1}{2}<(W_1W_2-\frac{1}{N_c}W)>_B,
 \ee
 where $W_1,W_2$ and $W$ are the Wilson loops in the fundamental
 representation along the closed contours $C_1, C_2$ and $C$ shown at
 the Figure. The averaging over background can be done using the
 cluster expansion method [6]. The average of two Wilson loops was
 calculated in [7],  and it was demonstrated that, assuming the
 existence of finite gluonic correlation length $T_g$, the
 generalized area law asymptotics may be obtained. For the Wilson
 loop average (12) one has (when both contours $C_1$ and $C_2$ are
 large):
  \be <{\cal{W}}>=\frac{N_c^2-1}{2} exp(-\sigma(S_1+S_2)),
 \ee
  where $\sigma$ is the string tension in the fundamental
 representation, and $S_1 $ and $S_2$ are the minimal  areas inside
 the contours $C_1$ and $C_2$. The area law (13) holds for
 practically all the reasonable configurations in the $q\bar{q}g$
 system, apart from the special case of the contours $C_1$ and $C_2$
 embedded into the same plane, when the expression (13) is replaced
 by
  \be <{\cal{W}}>=\frac{N_c^2-1}{2}
 exp(-\sigma(S_1-S_2)-\sigma^{adj}S_2),~~S_1>S_2
  \ee
   $\sigma^{adj}$ is
 the string tension in the adjoint representation.  Both regimes (13)
 and (14) match smoothly each other at the average distances between
 the contours $C_1$ and $C_2$ of order of correlation length $T_g$.
 Note that to derive the expression (13) we don't use the standard
 procedure [13] of appealing to the limit $N_c\to\infty$ and replacing
 the adjoint line in (10) by the double fundamental one. In actual
 calculations we, however, assume the regime (13) to be valid
 everywhere in the $q\bar{q}g$ configuration space, arriving in such
 a way to the picture of the $q\bar{q}g$ hybrid as a system of a
 gluon with two minimal strings attached, each having a quark (or
 antiquark) at  the end.

 To reduce  the four--dimensional dynamics in (9) to the
 three--dimensional one we follow the procedure  outlined in [8],
 using the parametrization for which
 $$z_{\mu}=(\tau,\vec{r}_q),~~
 \bar{z}_{\mu}=(\tau,\vec{r}_{\bar{q}}),~~
 Z_{\mu}=(\tau,\vec{r}_g),
 $$
 and introducing new dynamical variables
 $$
 \mu_1(\tau)=\frac{T}{2s}\dot{z}_0(\tau),~~
 \mu_2(\tau)=\frac{T}{2\bar{s}}\dot{\bar{z}}_0(\tau),~~
 \mu_3(\tau)=\frac{T}{2{S}}\dot{{Z}}_0(\tau)
 $$
   with $0\leq\tau\leq T$. The representation (9) for the Green
   function is now rewritten as
   \be
   G=\int D\vec{r}_q D\vec{r}_{\bar{q} }D\vec{r}_g D{\mu_1}D{\mu_2}
   D{\mu_3} exp(-A),
   \ee
   with the effective action
   \be
   A=\int^T_0d\tau\{\frac{m_q^2}{2\mu_1}+\frac{m_{\bar{q}}^2}{2\mu_2}+
   \frac{\mu_1\dot{r}^2_q}{2}+\frac{\mu_2\dot{r}^2_{\bar{q}}}{2}
   + \frac{\mu_3\dot{r}^2_g}{2}+
   \ee
   $$
   +\sigma\int^1_0d\beta_1\sqrt{\dot{w}^2_1w_1^{'2}-(\dot{w}_1w'_1)^2}
   +\sigma\int^1_0d\beta_2\sqrt{\dot{w}^2_2w_2^{'2}-(\dot{w}_2w'_2)^2}\},
   $$
   where the surfaces $S_1$ and $S_2$ are parametrized by
   the coordinated $w_{i\mu}$, and
   $$\dot{w}_{i\mu}=\frac{\partial w_{i\mu}}{\partial\tau},~~
   w'_{i\mu}=\frac{\partial w_{i\mu}}{\partial\beta_i},~~i=1,2.$$
   Assuming the straight--line approximation for the minimal
   surfaces,
   $$
   w_{1\mu}(\tau,\beta_1)=z_{\mu}(\tau)\beta_1+Z_{\mu}(\tau)(1-\beta_1),~~
w_{2\mu}(\tau,\beta_2)=\bar{z}_{\mu}(\tau)\beta_2+Z_{\mu}(\tau)(1-\beta_2),
   $$
   we arrive to the Lagrangian
\be
   {\cal{L}}=
   \frac{m_q^2}{2\mu_1}+\frac{m_{\bar{q}}^2}{2\mu_2}+
   \frac{\mu_1\dot{r}^2_q}{2}+\frac{\mu_2\dot{r}^2_{\bar{q}}}{2}
   + \frac{\mu_3\dot{r}^2_g}{2}+
   \ee
   $$
   +\sigma\rho_1\int^1_0d\beta_1\sqrt{1+l^2_1}+
   \sigma\rho_2\int^1_0d\beta_2\sqrt{1+l^2_2},
   $$
   $$l^2_1=(\beta_1\dot{\vec{r}}_q+(1-\beta_1)\dot{\vec{r}}_g)^2-
   \frac{1}{\rho^2_1}(\beta_1(\dot{\vec{r}}_q\vec{\rho}_1)+(1-\beta_1)
   (\dot{\vec{r}}_g\vec{\rho}_1))^2,
   $$
   $$l^2_2=(\beta_2\dot{\vec{r}}_{\bar{q}}+(1-\beta_2)\dot{\vec{r}}_g)^2-
   \frac{1}{\rho^2_2}(\beta_2(\dot{\vec{r}}_{\bar{q}}\vec{\rho}_2)
   +(1-\beta_2)
   (\dot{\vec{r}}_g\vec{\rho}_2))^2,
   $$
   $$\vec{\rho}_1=\vec{r}_q-\vec{r}_g,~~\vec{\rho}_2
   =\vec{r}_{\bar{q}}-\vec{r}_g.
   $$
   The Lagrangian (17) is the straightforward  generalization of the
   Lagrangian obtained in [8] for the $q\bar{q}$ string.

   To deal with the square root terms in (17) the auxiliary field
   approach was used in [8]. Here we are interested only in the masses
   of lowest states, so we adopt another strategy, which allows
   simple variational treatment of the problem.

   First, the experience borrowed from the results of papers [8] tells
   us that for low values of relative angular momenta the square
   roots in (17) can be expanded up to the second order in the
   transverse velocities $\vec{l}_i$, and it was proved to be a good
   approximation even for the massless constituents. This
   approximation corresponds to the  potential--like regime with the
   linear potential $V=\sigma\rho_1+\sigma\rho_2$, while the terms
   $\sim l^2_i$ (string corrections) can be taken into account as a
   perturbation.  With this simplification the Lagrangian (17)
   becomes quadratic in velocities, the centre--of--mass motion is
   decoupled, and the Hamiltonian (in the Minkowsky space--time) can
   be easily obtained from (17):
   \be
   H_0=\frac{m^2_q}{2\mu_1}+\frac{m^2_{\bar{q}}}{2\mu_2}+
   \frac{\mu_1+\mu_2+\mu_3}{2}+\frac{p^2}{2\mu_p}+\frac{Q^2}{2\mu_Q}
   + \sigma\rho_1+\sigma\rho_2
   \ee
   where the Jacobi coordinates $\vec{r}$ and $\vec{\rho}$ and
   conjugated momenta $\bar{p}$ and $\vec{Q}$  are introduced:
    $$
   \vec{r}=\vec{r}_q-\vec{r}_{\bar{q}},~
   ~\vec{\rho}=\vec{r}_g-
   \frac{\mu_1\vec{r}_q+\mu_2\vec{r}_{\bar{q}}}{\mu_1+\mu_2}
   $$
   $$
   \mu_p=\frac{\mu_1\mu_2}{\mu_1+\mu_2},~~
   \mu_Q=\frac{\mu_3(\mu_1+\mu_2)}{\mu_1+\mu_2+\mu_3}
      $$
   $$
   \vec{\rho}_1=-\vec{\rho}+\frac{\mu_2}{\mu_1+\mu_2}\vec{r},~~
   \vec{\rho}_2=-\vec{\rho}-\frac{\mu_1}{\mu_1+\mu_2}\vec{r}.
   $$
   The Hamiltonian (18) contains the fields $\mu_i(\tau)$, and one is
   to integrate over these fields in the path integral
   representation. Instead we proceed in a way suggested in [14]: we
   find eigenvalues $\varepsilon_0(\mu_i)$ of the Hamiltonian $H_0$
   variationally, assuming $\mu_i$ independent of $\tau$, and then
   minimize $\varepsilon_0(\mu_i)$ with respect to $\mu_i$. Such
   procedure works nicely for the ground states, with  the accuracy
   about several percent, and, apart from being very simple
   technically, allows for the approximate solution of the problem of
   defining the physical transverse states.

   Indeed, we have neglected  the spin dependence in the
    gluon
   Green function (6). Thus we do not expect the eigenfunctions of
   the Hamiltonian (18) to be transverse
   and they are really not. On the  other hand, within the framework
   of the variational procedure described above we can impose the
   constraint
   \be
   \mu_3\Psi_0-\mu_3(\dot{\bar{r}}_g\vec{\psi})=0,
   \ee
    projecting out
   the physical hybrid state $\Psi_{\lambda}=(\Psi_0,\vec{\Psi})$ (
   we restore for the moment the gluon spin index $\lambda$ in the
   state (2)).  The constraint (19) is compatible with the projector
   $P_{\mu\lambda}$(eq.(7))(after averaging over background and
   introducing the variables $\mu_i$). In accordance with the
   condition (19) and neglecting string corrections we choose the
   physical states to be transverse with respect to the
   (three--dimentional) gluon momentum:
    \be
   \vec{p}_3\vec{\Psi}=0,~~\Psi_0=0,
    \ee
     so that the states contain
   electric or magnetic gluons :  \be \vec{\Psi}^e_j\sim
     \vec{Y}_{jjm}(\hat{Q}) \ee

   \be
   \vec{\Psi}^m_j\sim\sqrt{\frac{j+1}{2j+1}}\vec{Y}_{jj-1m}
   (\hat{Q})+\sqrt{\frac{j}{2j+1}}
   \vec{Y}_{jj+1m}(\hat{Q})
      \ee
   where $j$ is the total momentum in the gluonic subsystem. (See
   the paper [15], where the gluonium spectrum was calculated within
   the similar assumptions).

   We stress that the choice (20), though looking quite natural, is
   nothing but the variational ansatz, and, in principle, cannot
   substitute for rigorous treatment of the problem of transverse
   and longitudinal gluonic degrees of freedom.
   The latter should be done with inclusion of spin variables into
   the path integral representation for the gluon Green function.

     As the illustrative example of our approach we present the
     results for the masses of lowest hybrids with $u$ and $d$
     quarks, assuming the latters to be massless. The radial part
     of the three--body wave function was chosen to be of the
     simple cluster form $\sim
     exp(-\alpha\rho^2_1-\alpha\rho^2_2)$(up to the centrifugial
     barrier), and the angular momenta were taken to be the lowest
     compatible with the ansatz (21) or (22). So the ground state
     hybrid contains a 1$^+$ or $1^-$ gluon and a quark--antiquark
     pair  with relative angular momentum $l=0$. More elaborated
     calculations for massive quarks, more realistic trial wave
     function and with inclusion of string corrections and
     short--range Coulomb force are in progress now and will be
     reported elsewhere.

     The parity and charge conjugation of the hybrids are given by
     \be
     P=(-1)^{l+j},~~C=(-1)^{l+s+1}
     \ee
     for the states with electric gluon (21), and
     \be
     P=(-1)^{l+j+1},~~C=(-1)^{l+s+1}
     \ee
     for the states with magnetic gluon (22), where $s$ is  the total spin
     of quarks. So the possible quantum numbers for the ground state hybrids
     are
     \be
     J^{PC}=0^{\mp +},~~1^{\mp +},~~2^{\mp +},~~ 1^{\mp -}.
     \ee
     With the chosen trial function these states are degenerate, and we
     expect this degeneracy to be removed mainly by spin--dependent effects.
     For the string tension $\sigma=0.2 GeV^2$ the eigenenergy of the
     ground state is estimated to be $E_0=3.3 GeV$.

     Now we enter the rather delicate point of setting the absolute scale
     for the mass of a hybrid. It is  well--known that in the potential
     models the large and negative additive constant in the potential is
     needed to fit the meson spectra. The same holds for the lowest
     $q\bar{q}$'s in the suggested picture: the negative constant
     $C\approx -0.7\div -0.8$ GeV is needed to describe the $S-,P-$
     and $D-q\bar{q}$ levels. At present  there is no reasonable
     model for this constant term, and the best way would be to
     adjust it by some hybrid candidate mass. In principle, the
     constant may be attributed to the perimeter term in the  formula
     for Wilson loop, or to the hadronic shift (creation of
     additional hadronic loops due to the string breaking). For the
     $q\bar{q}g$ system with two strings the most naive
     recipe to account for such mechanisms is
     that the constant is twice as large as for the $q\bar{q}$
     system. Taking this as an educational guess we end up with the
     mass \be M_0=1.7-1.8 GeV \ee for the ground $q\bar{q}g$ state.

     The value (26) should not of course be taken seriously. However, it
     appears to be rather close to the mass range where the hybrid
     candidates are believed to be placed, and to the  value
     obtained in the flux tube model [3,4]. Ideologically, our approach
     and the flux tube model have common features:  both models
     account for the transverse motion of the $q\bar{q}$ string,
     though the idea is realized in different ways.  The most
     important difference is in quantum numbers:  there is no
     constituent gluon in the flux tube, so instead of (25) one has
     [3]
      \be
       J^{PC}=0^{\mp\pm},1^{\mp\pm},2^{\mp\pm},1^{\pm\pm}
        \ee
         It
     is not easy to distinguish between models using the existing
     data in light quark sector. Indeed, at present several hybrid
     candidates in the mass range 1.5-2 GeV are under discussion:
     $1^{-+}$ exotic signal from BNL experiment [16], the $0^{-+}$
     and $2^{-+}$ states seen by VES collaboration [17], the $\pi
     (1775)$ state [18] which may be $1^{-+}$, but $0^{-+}$ or
     $2^{-+}$ assignement is not excluded, and additional vector
     states seen in $e^+e^-$ annihilation around 1.5-1.7 GeV [19].
     The hybrid assignement is attached to these  states basing on
     the very clear signature for hybrid decays in the flux tube
     model [20]: the decay of a hybrid into two $S$--wave mesons is
     suppressed, and it is really the case for the listed states.
     Nevertheless, just the same signature exists for a hybrid with
     electric constituent gluon [21,22], that means that the
     $P$--odd ground state hybrids (25) exhibit the same decay
     pattern.  So to tell one model from another one is to study the
     resonance activity in
     $0^{++},1^{++},2^{++},1^{+-}/0^{+-},1^{+-},2^{+-},1^{++}$
     channels around 1.5-2.0 GeV, a rather challenging experimental task.

     More promising is to use another selection rule, valid only for the
     decay of constituent hybrids [21,22]. There are two decay
     mechanisms.  One proceeds via string breaking, leads to the
     $q\bar{q}-q\bar{q}g$ final state  and is forbidden for the
     ground state because of phase space.  Another is due to the
     conversion of a gluon into $q\bar{q}$ pair. The constituent
     gluon behaves as the massive particle,  transferring its
     spin to the created pair, and it means that not only the total
     angular momentum but also the total spin of constituents
     conserved, giving rise to the set of selection rules for hybrid
     decays, similar to those obtained in [22].

     To conclude, we have demonstrated that the model for QCD
     quark--antiquark string can be extended to account for hybrid meson
     excitations.  The arising model of constituent glue appears to be
     compatible with the data on light quark meson spectroscopy.

We are grateful to Yu.A.Simonov for encouragement and many enlightening
discussions. We also thank K.G.Boreskov and A.Yu.Dubin for useful
comments.

The research described in this publication was made possible in part
by Grant N J77100 from the International Science Foundation and
Russian Government.

The support from Russian Fundamental Research  Foundation, Grant N
93-02-14937 is also acknowledged.

\newpage

{\Large Figure caption}\\

Wilson loop configuration corresponding to the propagation of the hybrid state.

      \end{document}